\documentclass[manuscript]{aastex}
 \usepackage{amsmath}

\newcommand{\Kepler}{{\em Kepler}}
\newcommand{\corot}{{\em CoRoT}}
\newcommand{\numax}{\mbox{$\nu_{\rm max}$}}
\newcommand{\Dnu}{\mbox{$\Delta \nu$}}

\newcommand{\muHz}{\mbox{$\mu$Hz}}

\newcommand{\half}{{\textstyle\frac{1}{2}}}

\shorttitle{Modelling of HD 186355 with \it Kepler}
\shortauthors{C. Jiang et al.}

\begin{document}

\title{Modelling \Kepler\ observations of solar-like oscillations in the red-giant star HD 186355}

\author{
C.~Jiang,\altaffilmark{1}
B.~W.~Jiang,\altaffilmark{1}
J.~Christensen-Dalsgaard,\altaffilmark{2}
T.~R.~Bedding,\altaffilmark{3}
D.~Stello,\altaffilmark{3}
D.~Huber,\altaffilmark{3}
S.~Frandsen,\altaffilmark{2}
H.~Kjeldsen,\altaffilmark{2}
C.~Karoff,\altaffilmark{2}
B.~Mosser,\altaffilmark{4}
P.~Demarque,\altaffilmark{5}
M.~N.~Fanelli,\altaffilmark{6}
K.~Kinemuchi,\altaffilmark{6}
F.~Mullally\altaffilmark{7}
}
\altaffiltext{1}{Department of Astronomy, Beijing Normal University,
 Beijing, 100875, China; \mbox{jiangchen@mail.bnu.edu.cn}}

\altaffiltext{2}{Department of Physics and Astronomy, Aarhus University, DK-8000 Aarhus C, Denmark}

\altaffiltext{3}{Sydney Institute for Astronomy (SIfA), School of Physics, University of Sydney, NSW 2006, Australia}

\altaffiltext{4}{LESIA, CNRS, Universit{\'e} Pierre et Marie Curie, Universit{\'e} Denis Diderot,
Observatoire de Paris, 92195 Meudon, France}

\altaffiltext{5}{Department of Astronomy, Yale University, P.O. Box 208101, New Haven, CT 06520-8101, USA}

\altaffiltext{6}{Bay Area Environmental Research Inst. NASA Ames Research Center, Moffett Field, CA 94035}

\altaffiltext{7}{SETI Institute NASA Ames Research Center, Moffett Field, CA 94035}

\begin{abstract}
We have analysed oscillations of the red giant star HD 186355 observed by the NASA \Kepler\ satellite. The data consist of the first five quarters of science operations of \Kepler, which cover about 13 months. The high-precision time-series data allow us to accurately extract the oscillation frequencies from the power spectrum. We find the frequency of the maximum oscillation power, $\numax$, and the mean large frequency separation, $\Dnu$, are around 106 and 9.4 \muHz\ respectively. A regular pattern of radial and non-radial oscillation modes is identified by stacking the power spectra in an \'{e}chelle diagram. We use the scaling relations of \Dnu\ and \numax\ to estimate the preliminary asteroseismic mass, which is confirmed with the modelling result ({\it M} = 1.45 $\pm$ 0.05 $M_{\sun}$) using the Yale Rotating stellar Evolution Code (YREC7). In addition, we constrain the effective temperature, luminosity and radius from comparisons between observational constraints and models. A number of mixed $l = 1$ modes are also detected and taken into account in our model comparisons. We find a mean observational period spacing for these mixed modes of about 58 s, suggesting that this red giant branch star is in the shell hydrogen-burning phase.
\end{abstract}


\section{Introduction}
Studying solar-like oscillations provides a powerful method to probe the interiors of stars \citep{chr04}. Solar-like oscillations are expected in low-mass main-sequence stars cooler than the red edge of the classical instability strip in the HR diagram \citep{chr82, chr83, hou99}, as well as in more evolved red giants which represent the future of our own Sun \citep{dzi01,dup09}. It is thought that turbulent convective motions near the surface excite the oscillations stochastically.

 Asteroseismology of red giants has developed rapidly. It began with several detections of solar-like oscillations in G and K-type giants based on ground-based observations in both radial velocity \citep{fra02, der06} and photometry \citep{ste07} and on space-based photometry detections observed by the {\em Hubble Space Telescope} ({\em HST}; \citealt{eng96,gil08,ste09}), {\em Wide Field Infrared Explorer} ({\em WIRE}; \citealt{buz00,ret03,ste08}), {\em Solar Mass Ejection Imager} ({\em SMEI}; \citealt{tar07}), and {\em Microvariability and Oscillations of Stars} ({\em MOST};  \citealt{bar07,Kal08a,Kal08b}). The oscillation periods in red giants range from hours to days. Ground-based observations usually suffer from interruptions and aliasing which complicate the measurement of oscillations. On the other hand, observations from space can provide high signal-to-noise ratio (SNR) and continuous data sets from which we may extract the oscillation parameters accurately. The 150-day observations by the {\em Convection Rotation and planetary Transits satellite} (\corot) clearly detected radial and non-radial oscillations in the range 10--100 \muHz\ \citep{der09,hek09,car10}, which greatly increased the number of detected pulsating G and K giants and led to a huge breakthrough in the study of red giants. These observations were followed by even more impressive results by \Kepler\ (e.g. \citealt{bed10a,hub10,kal10,hek11a,hek11b,mau11,cha11}).

This paper presents observations and models of HD 186355 (HIP 96878, KIC 11618103), which is one of the brightest red giants in the \Kepler\ field (V = 7.95).

\section{Observations}\label{sc:obv}
The \Kepler\ Mission \citep{bor08,bor10} was successfully launched on
March 7, 2009. Its primary scientific goal of is to search for Earth-sized planets in or near the habitable zone and to determine how many stars have this kind of planets in our Milky Way. \Kepler\ is equipped with a 0.95-meter diameter telescope with an array of CCDs which continuously points to a large area of the sky in the constellations Cygnus and Lyra to detect the transits of the planets. Over the whole course of the mission (at least 3.5 years), the spacecraft will measure the variations in the brightness of more than 100,000 stars, which will be outstanding data for the study of asteroseismology. For many of these stars we can detect solar-like oscillations, which will allow us to investigate them in detail and obtain their fundamental properties, by using the techniques of asteroseismology \citep{chr07,bookas}.

We used the first five quarters of data of HD 186355, which covers a total of about 13 months. The raw long-cadence data (29.4 minutes sampling; Jenkins et al. 2010) were corrected by performing a point-to-point sigma clipping to remove the outliers. Additionally, a thermal drift was corrected by fitting a second-order polynomial to the affected parts of the time-series. From the parallax of 5.44 $\pm$ 0.63 mas \citep{van07} and using a bolometric correction for G5 giants of -0.34 from \cite{kal89}, we derived the luminosity of the star to be 24.0 $\pm$ 5.6 $L_{\sun}$. We take the effective temperature ($T_{\rm eff}$ = 4867 $\pm$ 150 K) from \Kepler\ Input Catalogue (KIC; Brown et al. 2011).

\section{Global oscillation analysis}\label{sc:osc}
Solar-like oscillations are usually high-order and low-degree p-modes. Their frequencies are regularly spaced, approximately following the asymptotic relation \citep{tas80,gou86}:
\begin{equation}
  \nu_{nl} \approx \Dnu (n + \half l + \epsilon) - l(l+1) D_0,
        \label{eq:asy}
\end{equation}
where $n$ is the radial order and $l$ is the angular degree. The quantity $\Dnu$ (large frequency separation) is approximately the inverse of the sound travel time across the star, while $\epsilon$ is sensitive to the surface layers and, for relatively unevolved stars, $D_0$ is sensitive to the sound speed gradient near the core. As the star evolves, the stellar envelope starts to expand and the p-mode frequencies gradually decrease while oscillations in the core driven by buoyancy (g-modes) shift to higher frequencies. This eventually leads to so-called "mixed modes". These are non-radial oscillation modes that have a mixed character, behaving like g-modes in the core and p-modes in the envelope, and shifting in frequency as they undergo the so-called {\em avoided crossings} \citep{osa75,aiz77}. For red giants, the asymptotic $l$ = 1 modes in particular depart from the relation due to many avoided crossings \citep{hub10,mos11}.

Our frequency analysis covers three basic steps that are performed on the power spectrum of the \Kepler\ light curve: fitting and correcting for the background; estimating the frequency of maximum power ($\numax$) and the large separation ($\Dnu$); and extracting individual frequencies ($\nu_{nl}$). In the following subsections, we describe the three analysis steps in detail.

\subsection{Modelling the Background and Determining \numax }\label{sc:background}

The power spectrum shows a frequency-dependent background signal due to stellar activity, granulation and faculae which can be modelled by a sum of several Lorentzian-like functions \citep{har85}. In this paper, stellar activity, granulation and faculae were represented by modified Lorentzian-like functions, first introduced by \citet{karoff_phd}, which give a better fit to the background than a Harvey model with a constant slope of $-2$. This background model has a shallower slope at low frequencies and a steeper slope at higher frequencies, corresponding to stellar activity and granulation, respectively. The power excess hump from stellar oscillations is approximately Gaussian, so the complete spectrum was modeled by:
\begin{equation}
\begin{aligned}
P(\nu) =P_{n} + \sum_{i=1}^3 \frac{4\sigma_{i}^2\tau_{i}}{1+(2\pi\nu\tau_{i})^{2}+(2\pi\nu\tau_{i})^{4}}
 + P_\mathrm{g}\exp{\left(\frac{-(\numax-\nu)^{2}}{2\sigma_\mathrm{g}^{2}}\right)},
\end{aligned}
\label{eq:background}
\end{equation}
where $P_{n}$ corresponds to the white noise component, $\sigma_{i}$ is the rms intensity of the granules and $\tau_{i}$ is the characteristic time scale of granulation. For the Gaussian term, the parameters $P_\mathrm{g}$, $\nu_\mathrm{max}$, and $\sigma_\mathrm{g}$ are the height, the central frequency, and the width of the power excess hump.

\begin{figure}
\epsscale{1.0}
\plotone{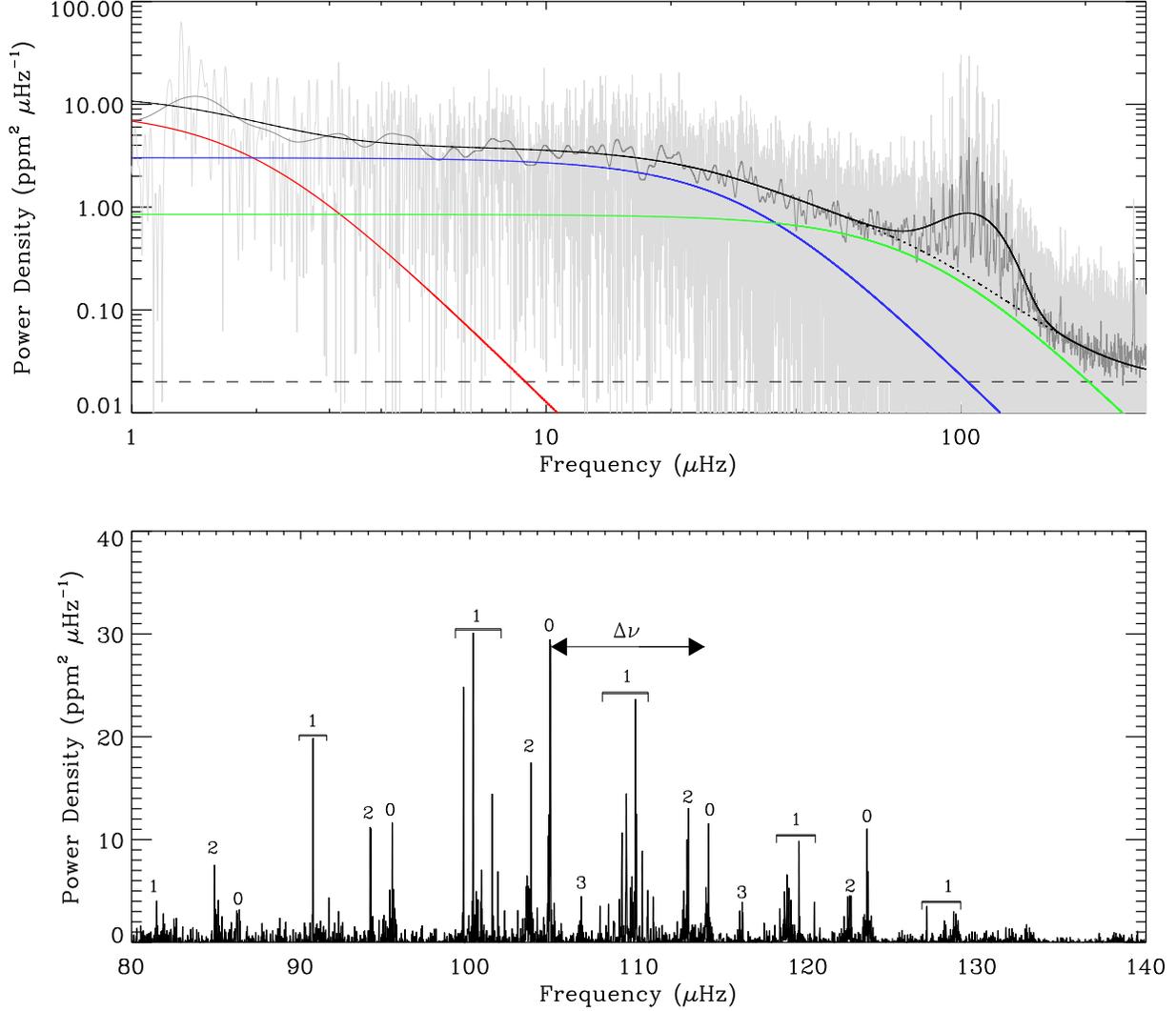}
\caption{Top panel: power density spectrum of the combined first five quarters of data (light-grey) and corresponding global model fit(black line). The dark-grey line is the smoothed (Gaussian with a FWHM of 0.5 $\muHz$) power density spectrum. The dotted line is the fitted background and the dashed line is the white noise. The red line shows the contribution from stellar activity, the blue from granulation and the green from faculae \citep{karoff_phd}. Bottom panel: background corrected power density spectrum in the range of the stellar oscillations. It is clear that \numax\ is around 106 \muHz\ and that peaks are regularly spaced with a large spacing of 9.37 $\muHz$. Numbers are degrees of the modes. The multiple peaks corresponding to modes of $l = 1$ are believed to be mixed modes \citep{bec11, bed11}.}
\label{fg:background}
\end{figure}

Fig.~\ref{fg:background} shows the power density spectrum of HD 186355, together with the fitted model using Eq.~(\ref{eq:background}). The three components of the background and the white noise were simultaneously fitted to a lightly smoothed power spectrum (Gaussian with FWHM of 0.5 $\muHz$) outside the region where the power excess hump is seen. The value of $\numax$ was obtained by fitting to a heavily smoothed power spectrum (Gaussian with FWHM of 3 $\Dnu$, where $\Dnu$ is estimated in Sect.~\ref{sc:freq}), giving 106.5 $\pm$ 0.3 $\muHz$. Finally, the background and the white noise were subtracted from the power density spectrum, leaving only the oscillation signal (lower panel of Fig.~\ref{fg:background}).

\subsection{Individual Frequencies} \label{sc:freq}

\begin{figure}
\epsscale{1.0}
\plotone{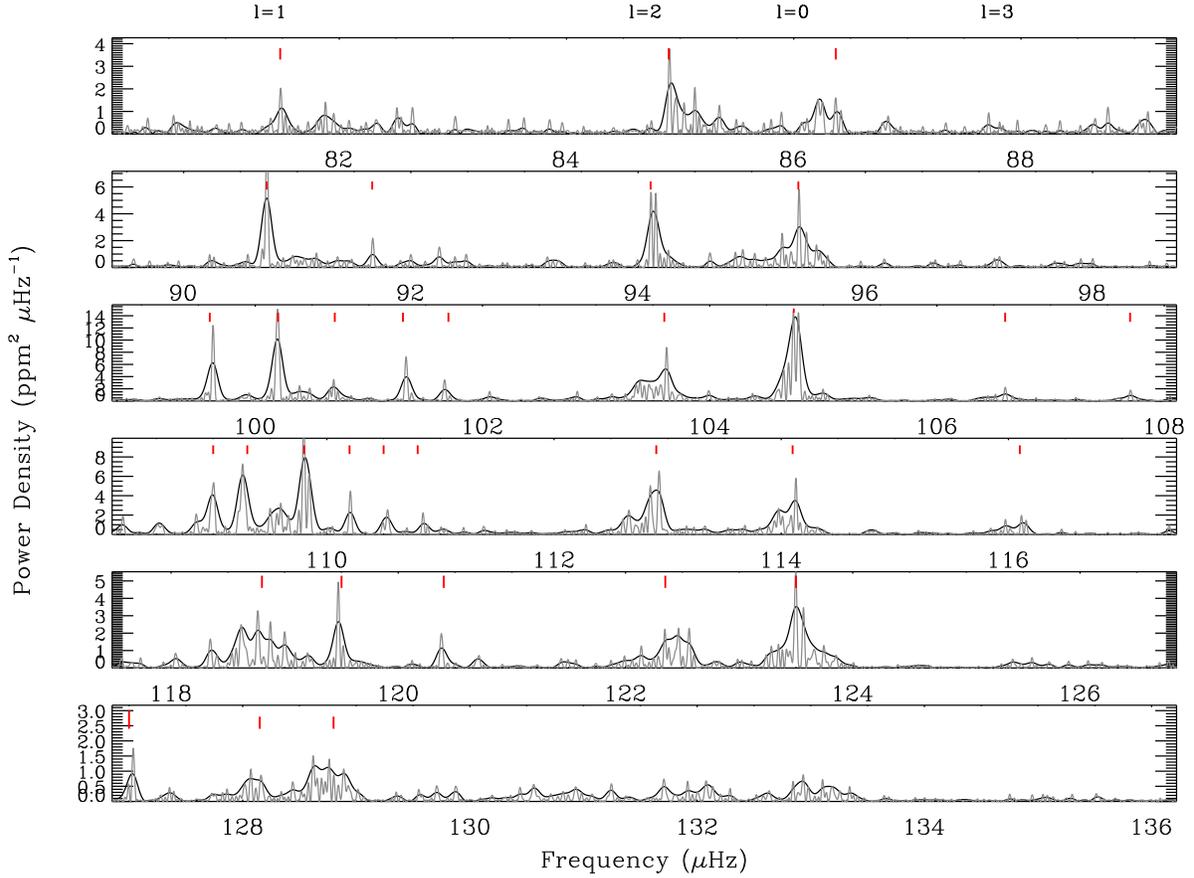}
\caption{The \'{e}chelle diagram of the smoothed power density spectrum (dark, FWHM of 0.1 $\muHz$) and the unsmoothed background corrected power density spectrum (grey, power divided by a factor of 2) divided into bins each $\Dnu$ wide. The red bars indicate 33 frequencies listed Table~\ref{tb:fre}. The peaks for the same degree almost line up. The offset from perfect alignment is caused by the variation of the large frequency spacing with frequency. Multiple peaks for $l = 1$ modes can be seen clearly.}
\label{fg:echps}
\end{figure}

The background-corrected power spectrum in Fig.~\ref{fg:background} shows the clear signature of solar-like oscillations: a regular series of peaks spaced by the large separation. We also see multiple peaks due to mixed $l=1$ modes (see also Sect.~\ref{sc:mod}, Beck et al. 2011, Bedding et al. 2011). The power spectrum is shown in \'{e}chelle format in Fig.~\ref{fg:echps}, both with and without smoothing. This diagram was made by dividing the power spectrum into six segments, each $\Dnu$ wide. We see that the peaks align vertically, allowing us to assign the $l$ values indicated on Figs.~\ref{fg:background} and~\ref{fg:echps}. We do not see an obvious signature of rotational splitting, and the effect of stellar rotation is not considered in this paper.

To extract the frequencies of individual oscillation modes, we used the software package Period04 \citep{len04}. This uses iterative sinewave fitting, which does a good job of extracting frequencies in cases such as this, where the individual modes are unresolved or barely resolved. The red lines in Fig.~\ref{fg:echps} show the frequencies of 33 extracted peaks with SNR greater than 3). These are listed in Table~\ref{tb:fre}, together with their amplitudes and uncertainties.

The uncertainties derived by Period04 are underestimates because they only consider the internal consistency of the parameters. We derived more realistic uncertainties (the second column in Table~\ref{tb:fre}) by means of Monte-Carlo simulations. The residual time-series $(t,y)$ were obtained by subtracting the sum of multiple sine functions of frequencies $(f_i)_{i=1,2,...,n}$ and the corresponding amplitudes $A_i$ and phases $\phi_i$ from the observed time-series $(t,x)$ as
\begin{equation}
y = x - \sum_{i=1,...,n} A_j\sin(2\pi{f_i}t + \phi{_i}).
\end{equation}
Then $|y|$ is regarded as the observational uncertainty in~$x$. We constructed 100 simulated time-series $z$, which have the same residuals as the observed time-series. The 100 simulated time-series were fitted with the sum of multiple sine functions by taking $(f_i,A_i,\phi_i)_{i=1,2,...,n}$ as initial values according to least-squares algorithm. Hence 100 sets of new $(f_i,A_i,\phi_i)_{i=1,2,...,n}$ were obtained. The standard deviations of each parameter of $(f_i,A_i,\phi_i)_{i=1,2,...,n}$ were then calculated, which were adopted as the uncertainty estimates of the parameters.

We also calculated the mean large frequency separation $\Dnu$ by performing a linear fit to the five $l = 0$ modes. Each data point was weighted by the uncertainty of the frequency listed in Table~\ref{tb:fre}. Frequencies for $l = 0$ modes are the most suitable ones for this calculation because they are not affected by the mixing with g modes. The slope of the fitted line gave the large separation to be $\Dnu = 9.37 \pm 0.03\,\mu$Hz.

\section{Modelling}\label{sc:mod}
The common way to estimate the fundamental properties is to compare calculated model parameters with the observational constraints. We employed the Yale rotating stellar evolution code (YREC; Demarque et al. 2008) for stellar evolution modelling computations, and the non-radial and non-adiabatic stellar pulsation programme JIG developed by \cite{gue94} for frequency calculations. YREC can evolve our models up to the tip of the red giant branch, which is adequate for HD 186355. The input physics of the current YREC version (YREC7) included the latest OPAL opacity tables \citep{igl96}, OPAL equation of state \citep{rog02} and NACRE reaction rates \citep{ang99}. At low temperatures, opacities are obtained from \cite{fer05}. Convection is treated under the assumption of {\em mixing length theory} \citep{boh58}. We did not take rotation, diffusion or convective overshoot into consideration in our calculation.

There are several main inputs in YREC7---mass, $\alpha_{ml}$ (to determine the mixing-length $l_{ml} = \alpha_{ml}H_p$, where $H_p$ is pressure scale height), hydrogen abundance ($X$) and heavy-element abundance ($Z$). The best models are searched among those grids after being compared with observational constraints. For our models, $\alpha_{ml}$ and $X$ were fixed to the solar values of 1.8 and 0.72, respectively. The value of $Z$ was varied within a certain range, usually from 0.005 to 0.025 with a step of 0.002, but it changes for models with different masses.

Our initial estimate for the mass was made using scaling relations. \cite{bro91} proposed a scaling relation that can be used to predict \numax\ by scaling from the solar case:
\begin{equation}
\frac{\nu_\mathrm{max}}{\nu_\mathrm{max,\sun}} \approx \left( \frac{M}{M_\sun} \right) \left(\frac{R}{R_\sun}\right) ^{-2} \left( \frac{T_\mathrm{eff}}{T_\mathrm{eff,\sun}}\right)^{-1/2}.
\end{equation}
This relation gives a very good estimate for \numax\ for less evolved stars \citep{bed03}, while \cite{ste08} have shown that it holds also for stars on the giant branch, although with larger uncertainties. \cite{hans95} give the scaling relation to predict \Dnu\:
\begin{equation}
\frac{\Dnu}{\Dnu_\sun} \approx \left( \frac{M}{M_\sun} \right)^{1/2}\left(\frac{R}{R_\sun}\right) ^{-3/2}.
\end{equation}
Knowing \numax, \Dnu\ and $T_{\rm eff}$, the stellar mass is estimated by:
\begin{equation}
\frac{M}{M_\sun} \approx \left( \frac{\Dnu}{\Dnu_\sun} \right)^{-4} \left( \frac{\numax}{\nu_\mathrm{max,\sun}} \right)^3  \left( \frac{T_\mathrm{eff}}{T_\mathrm{eff,\sun}}\right)^{3/2}.
\end{equation}
Using \numax\ and \Dnu\ from Sec.~\ref{sc:osc} and $T_\mathrm{eff}$ from KIC, the stellar mass is estimated as 1.41 $\pm$ 0.14 $M_\sun$. Therefore, the initial masses of our models were chosen to be within the range of 1.25 to 1.55 $M_\sun$ with a step of 0.01 $M_\sun$.

We looked for models for which the parameters are located inside the 1-$\sigma$ error box confined by the uncertainties of observational results in the H-R diagram. For these sets of modelling parameters, we used a fine resolution for $Z$ (in steps of 0.001) in order to find the best models. Some models with larger masses were also calculated, with a bigger mass step of 0.1 $M_\sun$, in an attempt to search for models in a large range, because the scaling relations and hence the estimated mass are not so reliable for the giant branch stars.

\begin{figure}
\epsscale{0.8}
\plotone{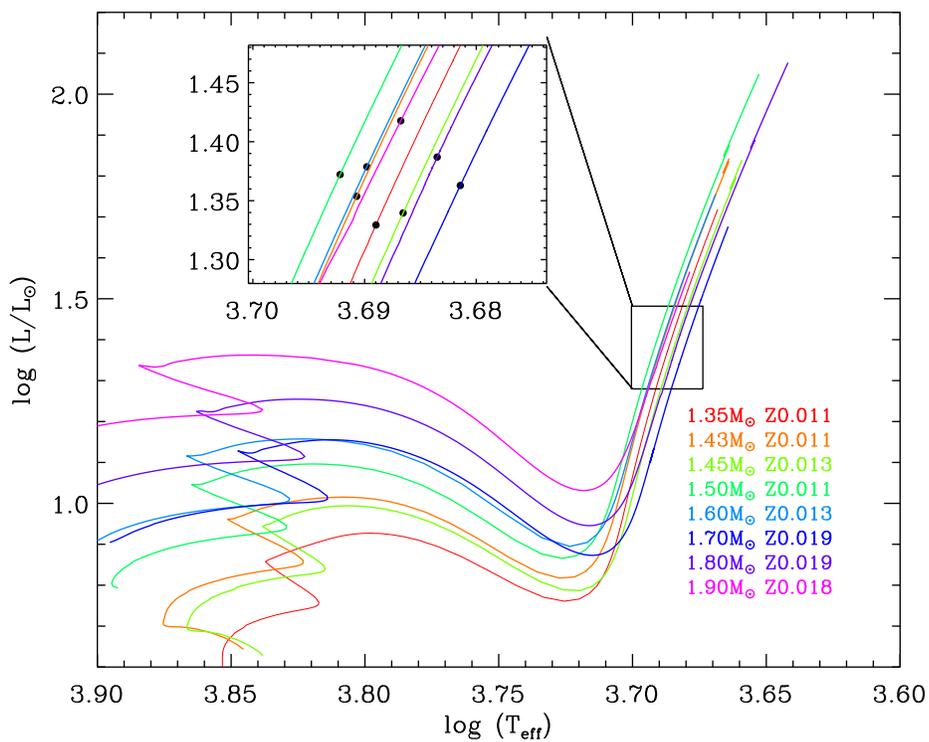}
\caption{Evolutionary tracks for eight models with different initial masses and heavy element abundances indicated by different colours listed in Table~\ref{tb:chi}. The rectangle is the 1$\sigma$ error box for the observational constraints, and dots are models which have a large frequency spacing close to 9.37 $\muHz$.}
\label{fg:track}
\end{figure}

Fig.~\ref{fg:track} shows several evolutionary tracks of models having different input parameters. The rectangle is the 1-$\sigma$ error box whose center corresponds to the observed stellar properties, from which we can see that HD 186355 is on the ascending giant branch, in the shell hydrogen-burning phase. Those models for which the parameters are within the error box and the mean large frequency separations are around 9.37 \muHz\ (within 0.03 \muHz) are indicated by dots. Different evolutionary tracks may pass through the same position in the H-R diagram by tuning the inputs. For example, a decrease of mass can be compensated by a decrease of hydrogen and heavy elements abundances to obtain the same position. Taking variations of the mixing-length into consideration, which move the tracks horizontally but have almost no influence on the luminosity, makes it even more complex to look for models. However, it is beyond the scope of this paper to consider the effects of varying the mixing-length and hydrogen abundance. We performed a $\chi^2$ minimization to find the best models. The definition of the $\chi^2$ function was based on two observed stellar parameters (luminosity and $T_\mathrm{eff}$) and on the individual frequencies, as follows:
\begin{equation}
\begin{aligned}
\chi^2 = \left(\frac{T_{\rm eff} - T_{\rm eff}^\prime}{150   \rm{K}}\right)^2+\left(\frac{\log L/L_\sun - \log L ^\prime/L_\sun}{0.11}\right)^2
+\frac{1}{N}\sum^N_{i=1} \left(\frac{\nu_i-\nu_i^\prime}{\sigma_i}\right)^2,
\end{aligned}
\label{eq:chi}
\end{equation}
where terms with primes are observed values, $N$ is the number of observed frequencies and $\sigma_i$ is the uncertainty of each frequency.  We did not find it necessary to apply an offset to the model frequencies to correct for near-surface effects \citep{hans08}.

\begin{figure}
\epsscale{1.0}
\plotone{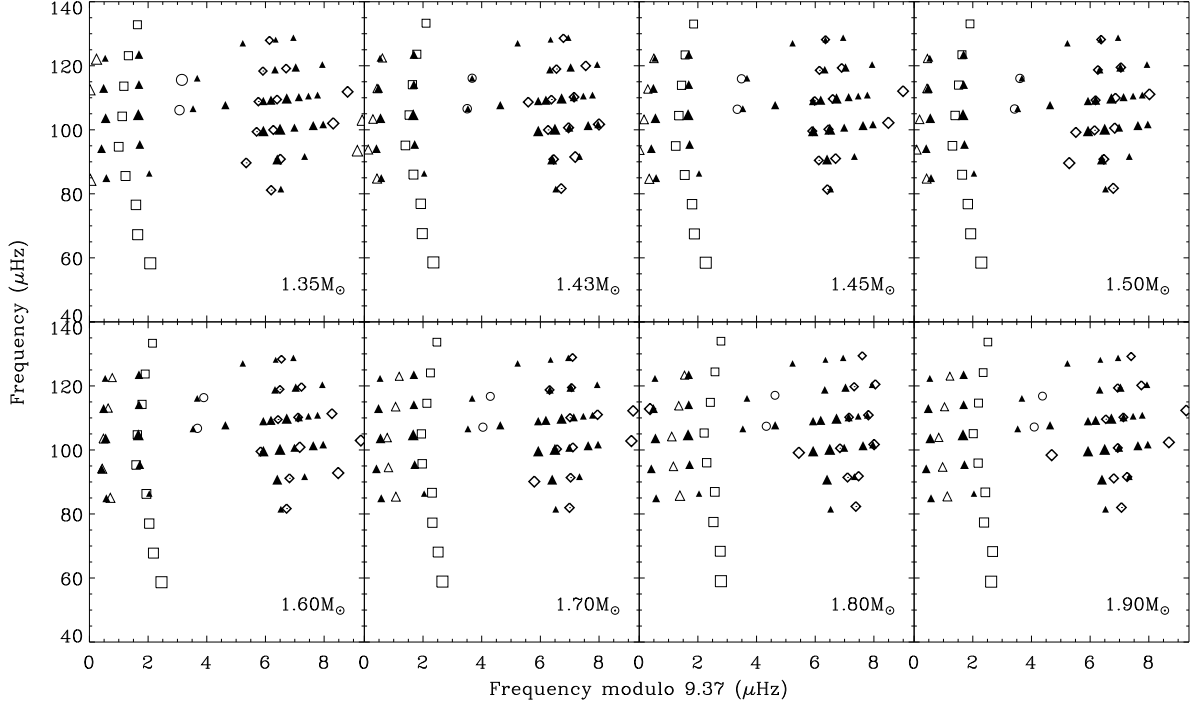}
\caption{\'{E}chelle diagrams for models plotted in Fig.~\ref{fg:track}. Squares, diamonds, triangles and circles are used for modes of degree $l = 0, 1, 2$, and 3, respectively. Observed frequencies are indicated by filled triangles. For theoretical frequencies, only those with a corresponding observed mode are shown (except for $l = 0$ modes) and symbol size indicates expected amplitude of each mode, which is scaled from mode inertia. For the observed ones, the size is scaled from the amplitude listed in Table~\ref{tb:fre}. Models with masses around 1.43 $M_{\sun}$ reproduce the observed frequencies better than those with higher masses. The differences between individual frequencies are included in the calculation of $\chi_1^2$. }
\label{fg:ech}
\end{figure}

\begin{figure}
\epsscale{0.8}
\plotone{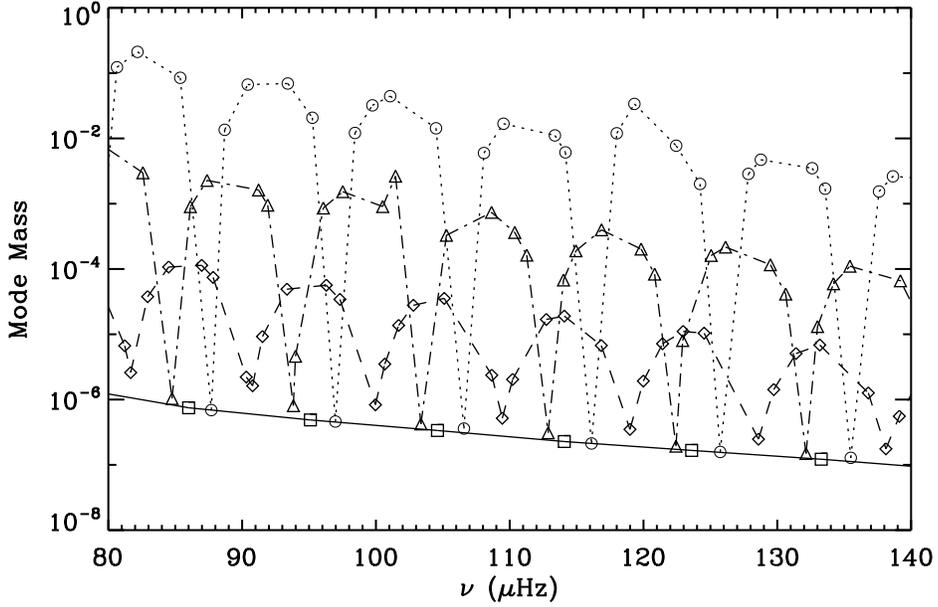}
\caption{Mode mass obtained by JIG9 for each degree (squares linked by solid line: $l = 0$, diamonds linked by dashed line: $l = 1$, triangles linked by dash dot line: $l = 2$, circles linked by dotted line: $l = 3$) versus frequency for the model with mass of 1.43 $M_\sun$ that has the smallest $\chi^2$. The dimensionless mode inertia is the ratio of mode mass to stellar mass. Those frequencies with minimal mode masses (hence minimal mode inertias) are plotted in the \'{e}chelle diagram in Fig.~\ref{fg:ech}.}
\label{fg:int}
\end{figure}

We began by fitting only one mode of each degree in each order. For $l$ = 1 we took the strongest peak in each order. The results based on these 17 observed frequencies are listed in Table~\ref{tb:chi}. Although tracks for models with masses larger than 1.60 $M_\sun$ also pass through the error box, their oscillation parameters differ greatly from the observations, which leads to bigger $\chi^2$. From Table~\ref{tb:chi} we can see the best model has a mass of 1.43 $M_\sun$ and $Z$ of 0.012. However, the difference between this and the 1.60 $M_\sun$ model is very small, which makes the latter one a candidate for the upper limit of the stellar mass estimation. In order to investigate this, we plot a series of \'{e}chelle diagrams in Fig.~\ref{fg:ech} to compare the theoretical and observed frequencies for each model shown in Table~\ref{tb:chi}. Only those frequencies having minima of mode inertia (see Fig.~\ref{fg:int}) are shown, because those modes will have the highest amplitudes at the stellar surface. Good agreement is found for 1.43 $M_\sun$, and models with larger masses do not reproduce the observed frequencies as well. However, the 1.60 $M_\sun$ model is an exception which produces a rather good match to observed frequencies for $l = 0$ and 2 modes. The location of the 1.60 $M_\sun$ model is close to the center of error box in HR diagram which, combined with the relatively good fit to the $l$ = 0 and 2, leads to a smaller $\chi^2$ result than models with higher masses. After being compared with the theoretical modes in \'{e}chelle diagrams, the degrees of observed modes (see Fig.~\ref{fg:echps}) are confirmed, including two modes with $l = 3$.

As mentioned in Sec.~\ref{sc:freq}, we found multiple oscillation peaks per order for $l = 1$ (see Table~\ref{tb:fre}) that we interpret as mixed modes. As discussed by \cite{bec11} and \cite{bed11}, it is impossible to observe some mixed modes (g-dominated mixed modes) because they have very high inertias. However, other mixed modes act more like p modes (p-dominated mixed modes), having a lower inertia than the g-dominated mixed modes and hence larger amplitude, which makes them observable. \cite{tas80} and \cite{mig08} have shown that pure g modes are equally spaced in period. P-dominated mixed modes are only approximately equally spaced in period. Measuring the period spacings for these observed mixed modes allows us to probe the cores of red giant stars. \cite{bec11} have detected mixed modes in a red giant star with \Kepler\ data and measured their period spacing. Subsequently, \cite{bed11} have found a way to distinguish between hydrogen-burning and helium-burning red giants by using their different period spacings. They found that hydrogen-shell burning stars have observed period spacings mostly around 50 s, while stars with helium-burning cores have observed period spacings of about 100 to 300 s. The observed mean period spacing of HD 186355 is 58 $\pm$ 4 s which we measured by means of power spectrum of the power spectrum method \citep{bed11}. This agrees with the value of 56 s for this star found by \cite{bed11}, and confirms that HD 186355 is still in the shell hydrogen-burning phase. This value also agrees with our models. We note that measuring the period spacings may also provide a method to determine the size of the convective core of those helium-burning red giants \citep{chr11}.

To make use of this extra information, we took frequencies of those $l = 1$ mixed modes with relatively low theoretical mode inertias into calculation. About 12 $l =1$ modes were used for each model, giving a total of around 24 frequencies. These produced the values labelled $\chi^2_1$ in Table~\ref{tb:chi}. Again the best model is the one with 1.43 $M_\sun$, but now the models with higher masses have large deviations between observed and theoretical oscillation frequencies. In particular, the 1.6 $M_\sun$ model is ruled out after this calculation. We search for models with $\chi^2_1$ smaller than 30 and determine the mass to be 1.45 $\pm$ 0.05 $M_\sun$.

\section{Conclusion}
We have analysed the time series data sets of the star HD 186355 from \Kepler\ to obtain its oscillation parameters. By using the scaling relations between $\Dnu$, $\numax$ and the stellar effective temperature $T_{\rm eff}$ we estimated the stellar mass as 1.41 $\pm$ 0.14 $M_{\sun}$. In order to determine the stellar global properties more accurately, we computed a set of models to compare with the observational constraints. The best model was found having a mass of 1.43 $M_\sun$, which agrees with the scaling value, and $Z$ of 0.012, and the stellar mass is constrained to be 1.45 $\pm$ 0.05 $M_\sun$. Furthermore, parameters such as age, effective temperature, luminosity and radius are also determined after comparison (see model with mass of 1.43 $M_\sun$ in Table~\ref{tb:chi}). We also obtain the observed mean period spacing of $l = 1$ modes with a value of 58 $\pm$ 4 s. From the modelled evolutionary track of HD 186355, we know it is in the shell hydrogen-burning phase and on the ascending giant branch, which is consistent with the results of \cite{bed11} on the mean period spacings of mixed modes for red giants.

\section{Acknowledgements}
The authors acknowledge the \Kepler\ Science Team for their work to provide us with these great data. Funding for the \Kepler\ Discovery mission is provided by NASA's Science Mission Directorate. This work is supported by China's NSFC through the project 10973004, China 973 Program 2007CB815406, and the Fundamental Research Funds for the Central Universities.

\clearpage
\begin{deluxetable}{ccccc}
\tabletypesize{\footnotesize} \tablecaption{Frequencies extracted by Period04 \label{tb:fre}} \tablewidth{0pt}
\tablehead{\colhead{\small{$l$}} &\colhead{\small{Frequency}}&\colhead{\small{Freq. sigma}} &\colhead{\small{Amplitude}} &\colhead{\small{SNR}}\\
\colhead{}&\colhead{\small{(\muHz)}}&\colhead{\small{(\muHz)}}&\colhead{\small{(ppm)}}&\colhead{}}
\startdata
\small1 &\small81.5 &\small0.15  &\small0.37 &\small3.1 \\
\small2 &\small84.9 &\small0.31  &\small0.50 &\small4.2 \\
\small0 &\small86.4 &\small0.56  &\small0.33 &\small3.0 \\
\small1 &\small90.7 &\small0.26  &\small0.79 &\small6.6 \\
\small1 &\small91.7 &\small0.25  &\small0.38 &\small3.1 \\
\small2 &\small94.1 &\small0.19  &\small0.61 &\small4.4 \\
\small0 &\small95.4 &\small0.09  &\small0.61 &\small4.8 \\
\small1 &\small99.6 &\small0.11  &\small0.88 &\small6.8 \\
\small1 &\small100.2 &\small0.39 &\small0.97 &\small7.1 \\
\small1 &\small100.7 &\small0.30 &\small0.55 &\small4.2 \\
\small1 &\small101.3 &\small0.06 &\small0.67  &\small5.9 \\
\small1 &\small101.7 &\small0.22 &\small0.47  &\small4.5 \\
\small2 &\small103.6 &\small0.21 &\small0.75  &\small6.8 \\
\small0 &\small104.7 &\small0.12 &\small0.96  &\small6.6 \\
\small3 &\small106.6 &\small0.66 &\small0.38 &\small3.2 \\
\small1 &\small107.7 &\small0.47 &\small0.54 &\small3.2 \\
\small1 &\small109.0 &\small0.86 &\small0.56  &\small3.7 \\
\small1 &\small109.3 &\small0.35 &\small0.67  &\small5.8 \\
\small1 &\small109.8 &\small0.58 &\small0.85  &\small6.7 \\
\small1 &\small110.2 &\small0.50 &\small0.55  &\small4.3 \\
\small1 &\small110.5 &\small0.29 &\small0.40  &\small3.0 \\
\small1 &\small110.8 &\small0.04 &\small0.38  &\small3.0 \\
\small2 &\small112.9 &\small0.21 &\small0.65  &\small5.7 \\
\small0 &\small114.1 &\small0.25 &\small0.60  &\small4.6 \\
\small3 &\small116.1 &\small0.25 &\small0.36 &\small3.0 \\
\small1 &\small118.8 &\small0.13 &\small0.45  &\small4.3 \\
\small1 &\small119.5 &\small0.46 &\small0.56  &\small4.6 \\
\small1 &\small120.4 &\small0.24 &\small0.36  &\small3.1 \\
\small2 &\small122.3 &\small0.12 &\small0.38  &\small3.2 \\
\small0 &\small123.5 &\small0.66 &\small0.59  &\small4.5 \\
\small1 &\small127.0 &\small0.44 &\small0.33  &\small3.1 \\
\small1 &\small128.2 &\small0.11 &\small0.28  &\small3.0 \\
\small1 &\small128.8 &\small0.08 &\small0.33  &\small3.0
\enddata
\end{deluxetable}


\clearpage
\begin{deluxetable}{cccccccccc}\rotate
\tablecolumns{10}
\tabletypesize{\scriptsize}
\tablecaption{Modelling results: fundamental properties and $\chi^2$ for models with different masses. The expression of $\chi^2$ is given by Eq.~(\ref{eq:chi}), while $\chi^2_1$ takes frequencies of mixed modes with low mode inertia into consideration. \label{tb:chi}} \tablewidth{0pt}
\tablehead{\colhead{\small{$M$/M$_{\sun}$}}&\colhead{\small{$Z$}}&
\colhead{\small{Age}}&\colhead{\small{T$_\mathrm{eff}$}}&
\colhead{\small{$L$/L$_{\sun}$}}&\colhead{\small{$R$/R$_{\sun}$}}&\colhead{\small{$\log g$}}&\colhead{\small{\Dnu}}&\colhead{\small{$\chi^2$}}&\colhead{\small{$\chi^2_1$}}\\
\colhead{}&\colhead{}&\colhead{\small{(Gyr)}}&\colhead{\small{(K)}}&\colhead{}&\colhead{}&\colhead{}
&\colhead{\small{(\muHz)}}&& }
\startdata
\small1.35  &\small0.011  &\small3.26 &\small4886 &\small21.35 &\small6.45 &\small2.95     &\small9.37     &\small18.4 &\small44.5\\
\small1.43  &\small0.012  &\small2.78 &\small4880 &\small21.98 &\small6.57 &\small2.96   &\small9.39     &\small2.1 &\small7.4\\
\small1.45  &\small0.013  &\small2.76 &\small4859 &\small21.86 &\small6.61 &\small2.96 &\small9.37  &\small4.4 &\small29.8\\
\small1.50  &\small0.011  &\small2.21 &\small4923 &\small23.56 &\small6.68 &\small2.96 &\small9.37  &\small2.8 &\small25.9\\
\small1.60  &\small0.013  &\small1.85 &\small4895 &\small23.92 &\small6.81 &\small2.98 &\small9.37  &\small4.0 &\small60.5\\
\small1.70  &\small0.019  &\small1.83 &\small4802 &\small23.06 &\small6.95 &\small2.98 &\small9.36  &\small26.1 &\small84.5\\
\small1.80  &\small0.019  &\small1.49 &\small4825 &\small24.38 &\small7.07 &\small3.00 &\small9.37  &\small66.9 &\small97.0\\
\small1.90  &\small0.018  &\small1.22 &\small4861 &\small26.16 &\small7.22 &\small3.00 &\small9.35  &\small36.6 &\small84.4\\
\hline
&&&&&&&&\\
\small{Observational} &&&\small4867 $\pm$ 150 &\small24.0 $\pm$ 5.6 &&&\small9.37 $\pm$ 0.03&\\
\small{ constraints}&&&&&&&\\
\enddata
\end{deluxetable}
\clearpage

\begin{thebibliography}{50}

\bibitem[Aerts et al.(2010)]{bookas} Aerts, C.,
Christensen-Dalsgaard, J.,
\& Kurtz, D.~W.\ 2010, Asteroseismology by C.~Aerts, J.~Christensen-Dalsgaard, and D.W.~Kurtz.~Springer, 2010

\bibitem[Angulo et al.(1999)]{ang99} Angulo, C., et al.\
1999, Nuclear Physics A, 656, 3

\bibitem[Aigrain et
al.(2004)]{aig04} Aigrain, S., Favata, F., \& Gilmore, G.\ 2004, \aap, 414, 1139

\bibitem[Aizenman et
al.(1977)]{aiz77} Aizenman, M., Smeyers, P., \& Weigert, A.\ 1977, \aap, 58, 41

\bibitem[Anklin et al.(1998)]{ank98} Anklin, M., Frohlich,
C., Wehrli, C.,
\& Finsterle, W.\ 1998, Structure and Dynamics of the Interior of the Sun and Sun-like Stars, 418, 91

\bibitem[{{Barban} {et~al.}(2007){Barban}, {Matthews}, {De Ridder}, {Baudin},
  {Kuschnig}, {Mazumdar}, {Samadi}, {Guenther}, {Moffat}, {Rucinski},
  {Sasselov}, {Walker}, \& {Weiss}}]{bar07}
{Barban}, C., {et~al.} 2007, \aap, 468, 1033

\bibitem[{Beck} {et~al.}(2011)]{bec11} Beck, P.~G., et al.\
2011, Science, 332, 205

\bibitem[Bedding
\& Kjeldsen(2003)]{bed03} Bedding, T.~R., \& Kjeldsen, H.\ 2003, PASA, 20, 203

\bibitem[Bedding et al.(2010\natexlab{a})]{bed10a} Bedding, T.~R., et al.\
2010\natexlab{a}, \apjl, 713, L176

\bibitem[Bedding et al.(2010\natexlab{b})]{bed10b} Bedding, T.~R., et al.\
2010\natexlab{b}, \apj, 713, 935

\bibitem[Bedding et al.(2011)]{bed11} Bedding, T.~R., et al.\
2011, \nat, 471, 608

\bibitem[B{\"o}hm-Vitense(1958)]{boh58} B{\"o}hm-Vitense, E.\
1958, \zap, 46, 108

\bibitem[Borucki et al.(2008)]{bor08} Borucki, W., et al.\
2008, IAU Symposium, 249, 17

\bibitem[Borucki et al.(2010)]{bor10} Borucki, W.~J., et al.\
2010, Science, 327, 977

\bibitem[Brown et al.(1991)]{bro91} Brown, T.~M., Gilliland, R.~L., Noyes, R.~W.\ 1991, \apj, 368, 599

\bibitem[Brown et al.(2011)]{brown11} Brown, T.~M., Latham,
D.~W., Everett, M.~E., \& Esquerdo, G.~A.\ 2011, \aj, submitted

\bibitem[{{Buzasi} {et~al.}(2000){Buzasi}, {Catanzarite}, {Laher}, {Conrow},
  {Shupe}, {Gautier}, {Kreidl}, \& {Everett}}]{buz00}
{Buzasi}, D., {Catanzarite}, J., {Laher}, R., {Conrow}, T., {Shupe}, D.,
  {Gautier}, III, T.~N., {Kreidl}, T., \& {Everett}, D. 2000, \apjl, 532, L133

\bibitem[Carrier et
al.(2010)]{car10} Carrier, F., et al.\ 2010, \aap, 509, A73

\bibitem[Chaplin et al.(2010)]{cha10} Chaplin, W.~J., et al.\
2010, \apjl, 713, L169

\bibitem[Chaplin et al.(2011)]{cha11} Chaplin, W.~J., et al.\
2011, \apjl, 732, L5

\bibitem[Christensen-Dalsgaard(1982)]{chr82}
Christensen-Dalsgaard, J.\ 1982, Advances in Space Research, 2, 11

\bibitem[Christensen-Dalsgaard
\& Frandsen(1983)]{chr83} Christensen-Dalsgaard, J., \& Frandsen, S.\ 1983, \solphys, 82, 469

\bibitem[Christensen-Dalsgaard(2002)]{chr02}
Christensen-Dalsgaard, J.\ 2002, Reviews of Modern Physics, 74, 1073

\bibitem[Christensen-Dalsgaard(2004)]{chr04}
Christensen-Dalsgaard, J.\ 2004, \solphys, 220, 137

\bibitem[Christensen-Dalsgaard et al.(2007)]{chr07}
Christensen-Dalsgaard, J., Arentoft, T., Brown, T.~M., Gilliland, R.~L.,
Kjeldsen, H., Borucki, W.~J.,
\& Koch, D .\ 2007, Communications in Asteroseismology, 150, 350

\bibitem[Christensen-Dalsgaard(2011)]{chr11} Christensen-Dalsgaard, J.\ 2011, arXiv:1106.5946

\bibitem[De Ridder et
al.(2006)]{der06} De Ridder, J., Barban, C., Carrier, F., Mazumdar, A., Eggenberger, P., Aerts, C., Deruyter, S., \& Vanautgaerden, J.\ 2006, \aap, 448, 689

\bibitem[De Ridder et al.(2009)]{der09} De Ridder, J., et
al.\ 2009, \nat, 459, 398

\bibitem[Demarque et
al.(2008)]{dem08} Demarque, P., Guenther, D.~B., Li, L.~H., Mazumdar, A., \& Straka, C.~W.\ 2008, \apss, 316, 31

\bibitem[Di Mauro et al.(2011)]{mau11} Di Mauro, M. P., et al.\ 2011, \mnras, in press

\bibitem[Dupret et
al.(2009)]{dup09} Dupret, M.-A., et al.\ 2009, \aap, 506, 57

\bibitem[Dziembowski et al.(2001)]{dzi01} Dziembowski, W.~A.,
Gough, D.~O., Houdek, G., \& Sienkiewicz, R.\ 2001, \mnras, 328, 601


\bibitem[{{Edmonds} \& Gilliland(1996)}]{eng96}
{Edmonds}, P.~D., \& Gilliland, R.~L. 1996, ApJ, 464, L157

\bibitem[Frandsen et
al.(2002)]{fra02} Frandsen, S., et al.\ 2002, \aap, 394, L5

\bibitem[Ferguson et al.(2005)]{fer05} Ferguson, J.~W.,
Alexander, D.~R., Allard, F., Barman, T., Bodnarik, J.~G., Hauschildt,
P.~H., Heffner-Wong, A., \& Tamanai, A.\ 2005, \apj, 623, 585

\bibitem[{{Gilliland}(2008)}]{gil08}
{Gilliland}, R.~L. 2008, \aj, 136, 566

\bibitem[Gilliland et al.(2010)]{gil10} Gilliland, R.~L., et
al.\ 2010, \pasp, 122, 131

\bibitem[{{Gough}(1986)}]{gou86}
{Gough}, D.~O. 1986, in Hydrodynamic and Magnetodynamic Problems in the Sun and
  Stars, ed. Y.~{Osaki} (Uni. of Tokyo Press), 117

\bibitem[Gough(1990)]{gou90} Gough, D.~O.\ 1990, Progress of
Seismology of the Sun and Stars, 367, 283

\bibitem[Guenther(1994)]{gue94} Guenther, D.~B.\ 1994, \apj,
422, 400

\bibitem[Harvey(1985)]{har85} Harvey, J.\ 1985, Future
Missions in Solar, Heliospheric \& Space Plasma Physics, 235, 199

\bibitem[Hekker et
al.(2009)]{hek09} Hekker, S., et al.\ 2009, \aap, 506, 465

\bibitem[Hekker et al.(2011{\natexlab{a}})]{hek11a} Hekker, S., et al.\ 2011{\natexlab{a}}, \mnras, 414, 2594

\bibitem[Hekker et
al.(2011{\natexlab{b}})]{hek11b} Hekker, S., et al.\ 2011{\natexlab{b}}, \aap, 525, A131

\bibitem[Houdek et
al.(1999)]{hou99} Houdek, G., Balmforth, N.~J., Christensen-Dalsgaard, J., \& Gough, D.~O.\ 1999, \aap, 351, 582

\bibitem[Huber et al.(2010)]{hub10} Huber, D., et al.\ 2010,
\apj, 723, 1607

\bibitem[Iglesias
\& Rogers(1996)]{igl96} Iglesias, C.~A., \& Rogers, F.~J.\ 1996, \apj, 464, 943

\bibitem[Jenkins et al. (2010)]{jen10} Jenkins, J.~M., et al.\ 2010, \apjl, 713, L120

\bibitem[Kaler(1989)]{kal89} Kaler, J.~B.\ 1989, Cambridge: Cambridge University Press

\bibitem[{{Kallinger} {et~al.}(2008{\natexlab{a}}){Kallinger}, {Guenther},
  {Weiss}, {Hareter}, {Matthews}, {Kuschnig}, {Reegen}, {Walker}, {Rucinski},
  {Moffat}, \& {Sasselov}}]{Kal08a}
{Kallinger}, T., {et~al.} 2008{\natexlab{a}}, Commun. Asteroseismology, 153, 84

\bibitem[{{Kallinger} {et~al.}(2008{\natexlab{b}}){Kallinger}, {Guenther},
  {Matthews}, {Weiss}, {Huber}, {Kuschnig}, {Moffat}, {Rucinski}, \&
  {Sasselov}}]{Kal08b}
{Kallinger}, T., {et~al.} 2008{\natexlab{b}}, \aap, 478, 497

\bibitem[Kallinger et
al.(2010)]{kal10} Kallinger, T., et al.\ 2010, \aap, 522, A1

\bibitem[{{Karoff}(2008)}]{karoff_phd}{Karoff}, C. 2008, PhD thesis, Department of Physics and Astronomy, University of Aarhus

\bibitem[Kippenhahn
\& Weigert(1990)]{1990sse..book.....K} Kippenhahn, R., \& Weigert, A.\ 1990, Stellar Structure and Evolution, XVI, 468 pp.~192 figs..~ Springer-Verlag Berlin Heidelberg New York.~Also Astronomy and Astrophysics Library,

\bibitem[Kjeldsen
\& Bedding(1995)]{hans95} Kjeldsen, H., \& Bedding, T.~R.\ 1995, \aap, 293, 87

\bibitem[Kjeldsen et al.(2008)]{hans08} Kjeldsen, H., Bedding, T.~R., \& Christensen-Dalsgaard, J.\ 2008, \apjl,, 683, L175


\bibitem[Latham et al.(2005)]{lat05} Latham, D.~W., Brown,
T.~M., Monet, D.~G., Everett, M., Esquerdo, G.~A.,
\& Hergenrother, C.~W.\ 2005, Bulletin of the American Astronomical Society, 37, \#110.13

\bibitem[Lenz
\& Breger(2004)]{len04} Lenz, P., \& Breger, M.\ 2004, The A-Star Puzzle, 224, 786

\bibitem[Mazumdar(2005)]{maz05} Mazumdar, A.\ 2005, \aap, 441, 1079

\bibitem[Miglio et al.(2008)]{mig08} Miglio, A.,
Montalb{\'a}n, J., Eggenberger, P.,
\& Noels, A.\ 2008, Astronomische Nachrichten, 329, 529

\bibitem[Mosser et
al.(2010)]{mos10} Mosser, B., et al.\ 2010, \aap, 517, A22

\bibitem[Mosser et
al.(2011)]{mos11} Mosser, B., et al.\ 2011, \aap, 525,L9

\bibitem[Osaki(1975)]{osa75} Osaki, J.\ 1975, \pasj, 27, 237

\bibitem[{{Retter} {et~al.}(2003){Retter}, {Bedding}, {Buzasi}, {Kjeldsen}, \&
  {Kiss}}]{ret03}
{Retter}, A., {Bedding}, T.~R., {Buzasi}, D.~L., {Kjeldsen}, H., \& {Kiss},
  L.~L. 2003, \apjl, 591, L151

\bibitem[Rogers
\& Nayfonov(2002)]{rog02} Rogers, F.~J., \& Nayfonov, A.\ 2002, \apj, 576, 1064

\bibitem[{{Stello} {et~al.}(2007){Stello}, {Bruntt}, {Kjeldsen}, {Bedding},
  {Arentoft}, {Gilliland}, {Nuspl}, {Kim}, {Kang}, {Koo}, {Lee}, {Sterken},
  {Lee}, {Jensen}, {Jacob}, {Szab{\'o}}, {Frandsen}, {Csubry}, {Dind},
  {Bouzid}, {Dall}, \& {Kiss}}]{ste07}
{Stello}, D., {et~al.} 2007, \mnras, 377, 584

\bibitem[{{Stello} {et~al.}(2008){Stello}, {Bruntt}, {Preston}, \&
  {Buzasi}}]{ste08}
{Stello}, D., {Bruntt}, H., {Preston}, H., \& {Buzasi}, D. 2008, \apjl, 674,
  L53

\bibitem[{{Stello} \& {Gilliland}(2009)}]{ste09}
{Stello}, D., \& {Gilliland}, R.~L. 2009, \apj, 700, 949

\bibitem[{{Tarrant} {et~al.}(2007){Tarrant}, {Chaplin}, {Elsworth},
  {Spreckley}, \& {Stevens}}]{tar07}
{Tarrant}, N.~J., {Chaplin}, W.~J., {Elsworth}, Y., {Spreckley}, S.~A., \&
  {Stevens}, I.~R. 2007, \mnras, 382, L48

\bibitem[Tassoul(1980)]{tas80} Tassoul, M.\ 1980, \apjs, 43,
469

\bibitem[Vandakurov(1967)]{van67} Vandakurov, Y.~V.\ 1967,
\azh, 44, 786

\bibitem[van Leeuwen(2007)]{van07} van Leeuwen, F.\ 2007,
Astrophysics and Space Science Library, 350,


\end{thebibliography}
\end{document}